\documentclass[12pt]{article}
\usepackage{epsf}
\usepackage[intlimits]{amsmath}
\usepackage{graphicx}
\newcommand{\cd}{\makebox[0.08cm]{$\cdot$}}
\newcommand{\dd}{\text{d}}%

\title{\bf Spin zero particle propagator from a random walk in 3-D space}
\author{J.J. Dugne$^{a,}$\thanks{\textit{E-mail:} dugne@clermont.in2p3.fr}
\quad and \quad
V.A. Karmanov$^{b,}$\thanks{\textit{E-mails:} karmanov@sci.lebedev.ru,
karmanov@isn.in2p3.fr}\\
$^a${\small \em Laboratoire de Physique Corpusculaire,
Universit\'e Blaise-Pascal, CNRS/IN2P3}
\\{\small \em 24 avenue des Landais, F-63177 Aubi\`ere Cedex, France}\\
$^b${\small \em Lebedev Physical Institute, Leninsky Prospekt 53, 119991
Moscow, Russia}}

\date{}

\begin{document}
\bibliographystyle{unsrt}
\maketitle
\begin{abstract}
The  propagator of a spin zero particle in coordinate space is
derived supposing that the
particle propagates rectilinearly always at the speed of light and  changes its
direction in some random points due to a  scattering process. The
average path between two scatterings  is of the order of
the Compton length.
\end{abstract}


\section{Introduction}
Stochastic  problems in physics were extensively studied and this
approach has shown its great efficiency since a long time (see,
for instance, \cite{Chandrasekhar}). The attempts to understand,
from stochastical point of view, the quantum behaviour, have a
long history (see, e.g., \cite{Bohm52,Nelson84}).

The derivation of the  electron propagator in a 1-D space from a
random walk  assumption was carried out in the paper
\cite{Karmanov93}, where the motivation of this derivation is
explained in detail. It was supposed that electron moves with the
speed of light and randomly changes its direction of motion. The
average distance between the changes of direction was shown to be
exactly equal to the electron Compton length.  In a different but,
to some extent, analogous  approach, the electron propagator was
derived in  \cite{Riazanov58}.

In the present work we carry out the same calculations in a 3-D
space to see what assumptions have to be made to get  the correct
propagator of a massive zero spin particle.

\section{The propagator derivation}\label{prop}
Like  in the previous paper \cite{Karmanov93}, we suppose that the
particle  always moves in straight line at the speed of light.
 From time to time it encounters on its way something in the space  and
isotropically scatters with the amplitude $f$.  The process goes on until, at
the time $t$, the particle arrives in the space point $\vec{r}$. Integrating
over $n$  scattering points and taking the sum over $n$, we reproduce the spin
zero particle propagator.

We use the term "scattering" for convenience. Indeed, we do not localize the
particle between the different scatterings. Therefore one could equally give
another interpretation, like in the paper \cite{Kirilyuk00}, assuming the
re-creation of the  particle at the random time moments $t_i$, in  the points
separated from each other by the distances $r_i=ct_i$. Scatterings on
some quantum
vacuum fluctuations like in \cite{Nelson84} could also be invoked.
These issues are beyond  the scope of the present paper.

At present, we take our assumptions as postulates. It is instructive
to see, how the scattering
amplitude looks  in order to reproduce the propagator. Namely, we suppose
that after $i$-th scattering the wave function has the form:
\begin{equation}\label{rw1}
\psi(r_i)=\frac{f}{4\pi \ell_0^2 r_{i}},
\end{equation}
where $r_{i}$ is the distance from the $i$-th scattering center
and $\ell_0=\hbar/m c$ is the Compton length which will be our unit of length
throughout.

The factor $1/(4 \pi r_{i})$ in (1) gives the dependence on the
distance from the scattering center. It corresponds to standard
decreasing of a spherical wave. The factor $1/\ell_0^2$ is
introduced to make the amplitude $f$ dimensionless.
After each scattering the amplitude is  multiplied by $\psi(r_i)$
and integrated over $d^3r_i$, so with the factor $1/\ell_0^2$ in
(\ref{rw1}) the product $\psi(r_i) d^3r_i$ is dimensionless too.

We suppose also that $f=-1$, i.e. that our scattering process
mimics the resonance S-wave scattering with $\delta=\pi/2$.

Let us calculate  the amplitude $I_{n}$ over all the possible
paths consisting of straight intervals with $n-1$ changes of
direction. We denote by $\vec{r}$ the distance between the
extremal points in the propagator and $t$ is the necessary time to
the particle to pass along that distance.

As mentioned, with our assumptions we have $r_{i} = c t_{i}$, the path covered
between two changes of direction, $t_{i}$ the corresponding time interval. We
will take $c=1$ all along and so the Compton length will be $\ell_{0}=\hbar/m$
in this system of units.

The amplitude with $n-1$ changes of direction is represented in the form:
\begin{eqnarray}\label{rw2}
I_{n}&=&(-1)^{n-1}\int\,\delta^{3}\left(\vec{r}_{1} + \vec{r}_{2} +
\cdots +\vec{r}_{n}-\vec{r}\right)
\nonumber\\
&&\times\delta(r_{1} + r_{2} + \cdots + r_{n}-t)\,\frac{1}{4\pi r_1}
d^{3}r_{1}\psi(r_2)\,
d^{3}r_{2}\psi(r_3) \,
\cdots
\psi(r_n) d^{3}r_{n}
\end{eqnarray}
The function $\delta(r_{1} + r_{2} + \cdots + r_{n}-t)$ takes into
account the condition  that the sum of times $t_i=r_i/c$ gives the
total time $t$. It results from the function $\delta(t_{1} + t_{2}
+ \cdots + t_{n}-t)$ where, for $c=1$, we can change $t_{i}$ into
$r_i$.

The first scattering occurs in the point $\vec{r}_1$.
The particle arrives in this point
without any scattering. It is reflected by the factor
$\dfrac{1}{4\pi r_1}$ which does not contain the scattering amplitude $f$.
After the first scattering the particle is described by the wave function
$\psi(r_2)$, etc.

 The integration in the equation (\ref{rw2}) is carried out over
the points $\vec{r}_1, \vec{r}_2,\ldots$ and we can say nothing
about the particle on the path between these points.

It turns out that we only have to calculate $I_{2}$, the amplitude
with one change of direction, afterwards all the other amplitudes will be
deduced by iteration. We give the details below. First we start with:
\begin{equation}\label{rw3}
I_{2}=(-1)\int\, \delta^{3}\!\left({\vec{r}_{1} + \vec{r}_{2} -
\vec r}\right)\,\delta(r_{1}+r_{2}-t)\,\frac{d^{3}r_{1}}{4\pi\,r_{1}}\,
\frac{d^{3}r_{2}}{4\pi\,r_{2}\,{\ell_{0}}^{2}}
\end{equation}

After integration over  $d^3r_{1}$, we get:
\[
I_{2}\ =\frac{-1}{(4\pi\,{\ell_{0}})^2}
\int\,\delta\!\left(r_{1}+r_{2}-t\right)\,\frac{d^{3}r_{2}}
{r_{1}r_{2}}
\]
with $r_{1} = \left|\vec{r}-\vec{r}_{2}\right| =  \left(
r^{2}+r_{2}^{2}-2 r r_{2} \cos\theta_{2}\right)^{1/2}$.
\par\noindent

This integral is rewritten as:

$$
I_{2}=\frac{-1}{8\pi\,{\ell_{0}}^{2}}\,\int\,
\delta\left(r_{1}+r_{2} -t\right) \frac{r_{2}\,dr_{2}d z_2}{r_{1}}
= \frac{-1}{8\pi\,{\ell_{0}}^{2}}\,\int_0^{\infty} r_2
dr_2\int_{-1}^1\, \delta\left(f(z_2)\right) \frac{d z_2}{r_{1}}
$$
where we put $z_2 = \cos{\theta_{2}}$ and $f(z_{2}) = r_{2}-t+
r_{1}$; consequently
$$\frac{df(z_2)}{d z_2}\ =-\frac{r\,r_{2}}{r_{1}}.$$
The value  $z_{0}$ for which $f(z_0)=0 $ reads:
$$z_{0}\ =\frac{r^{2}+2t r_{2} -t^{2}}{2\,r\,r_{2}},$$
that gives $r_1=t-r_2$. The condition $-1 \leq z_0 \leq 1$
restricts the domain of $r_2$: $t-r < r_2 <t+r$. Integrating over
$dz_2$, we get:
$$
I_{2}=-\frac{1}{8\pi\,{\ell_{0}}^{2}}\,\frac{1}{r}\int_{t-r}^{t+r}
dr_2.
$$

In this way we obtain the following result:
\begin{equation}\label{rw4}
I_{2}\ =-\dfrac{1}{8\pi\,{\ell_{0}}^{\!\!2}}
\end{equation}
We can now look at the $I_{3}$ integral. It is given by :
\[
I_{3}=(-1)^2\int\,\delta^{3}\!\left(\vec{r}_{1} + \vec{r}_{2} -
(\vec{r} -\vec{r}_{3})\right)\,
\delta(r_{1} + r_{2} - (t - r_{3}))
\,\dfrac{d^{3}r_{1}}{4\pi\,r_{1}}
\,\dfrac{d^{3}r_{2}}{4\pi\,r_{2}\,{\ell_{0}}^{2}}
\,\dfrac{d^{3}r_{3}}{4\pi\,r_{3}\,{\ell_{0}}^{2}}
\]
So we can write this integral in a simpler form by using the former
$I_{2}$ result,
$$
I_{3}=-I_{2} \int
\frac{d^{3}r_{3}}{4\pi\,r_{3}\,{\ell_{0}}^{2}}\\
=-I_{2}\dfrac{1}{2\,\ell_{0}^{2}} \int r_{3}\,dr_{3}d z_3
$$
with  $z_3 = \cos{\theta_3} =
\dfrac{\vec{r}\cd\vec{r}_{3}}{rr_3}$. We introduce the variable:
$$
(\tau^{\prime})^{2} = (t-r_{3})^{2}-(\vec{r}-\vec{r}_{3})^{2} =
\tau^{2}-2r_{3} (t-r z_3)$$ with $\tau^{2} = t^{2}-r^{2}$. Then we
can replace the variable $r_3$ by ${\tau'}^2$ using the formula:
$r_{3} =\dfrac{(\tau ^{2}-{\tau^{\prime}}^{2})}{2(t-rz_3)}$ which
gives $d r_{3} = -\dfrac{d {\tau^{\prime}}^2}{2(t-rz_3)}$. The
integral $I_3$ is reduced to:
\[
I_{3}\ = -I_{2} \dfrac{1}{2 \ell_{0}^{2}} \int_{0}^{\tau^{2}}\dd
{\tau^{\prime}}^2\! \int _{-1}^{+1}d z_3\dfrac{(\tau
^{2}-{\tau^{\prime}}^{2})}{4\,(t-rz_3)^{2}}
\]
Taking into account that:
\[
\int _{-1}^{+1}\dfrac{\dd z}{(t-rz)^{2}} =\dfrac{2}{\tau ^{2}}
\]
we represent  the $I_{3}$ integral  in the form:
\[
I_{3}(\tau) = -I_{2}\dfrac{1}{\ell_{0}^{2}}  \int _{0}^{\tau^{2}}
\dfrac{(\tau ^{2}-{\tau^{\prime}}^2)}{4 \tau^{2}}\dd
{\tau^{\prime}}^{2}
\]
Where $I_2$ has the constant value given in the equation (\ref{rw4}), whereas $I_3$ depends on $\tau$.

The method can be similarly generalized to the integral of order
$n$ and gives:
\begin{equation}\label{rw5}
I_{n}(\tau)\ = -\dfrac{1}{\ell_{0}^{2}} \int_{0}^{\tau}
I_{n-1}\bigl({{\tau^{\prime}}}\bigr)\dfrac{(\tau
^{2}-{\tau^{\prime}}^{2})}{2 \tau^{2}}\tau^{\prime}\dd {\tau^{\prime}}
\end{equation}
with as usual   $\tau ^{2} =t^{2}-r^{2}$    and ${\tau^{\prime}}^{2}
= (t-r_{n})^{2}-(\vec r-\vec {r_{n}})^{2}$,
without forgetting the first integral:
\begin{equation}\label{rw5a}
I_{1}\ = \int \delta^{3}\!\left({\vec{r}_{1} -
\vec r}\right)\,\delta \left(r_{1}-t\right)
\frac{d^{3}r_{1}}{4\pi\,r_{1}} =
\frac{1}{4\pi r}
\delta\left(r-t\right)=\frac{1}{2\pi}\delta\left(\tau^2\right)
\end{equation}
This integral also contains the factor  $1/(4\pi r_{1})$
responsible for a decrease of the outgoing wave. But since the
first scattering contributes into $I_{2}$, the integral $I_{1}$
does not contain the scattering amplitude $f$.

The propagator of the particle will be obtained in summing all the
amplitudes corresponding to all the possible numbers of changes of
direction, i.e.
\[
K(\vec r,t) = I_{1}+I_{2}+I_{3}+\cdots+I_{n}+\cdots
\]
It is possible to do exactly this summation, the result being:
\[
K(\vec r,t) = \frac{1}{2\pi}\delta\left(\tau^2\right)
-\dfrac{1}{8\pi{\ell_{0}}^{2}}
\sum_{n=0}^{+\infty}\dfrac{(-1)^n{\left (\tau^{2}/{\ell_{0}}^{2}\right)}^{n}}
{(2^{n})^{2}(n+1)(n!)^{2}}
\]
Finally we obtain the propagator in a compact form using $J_{1}$ the Bessel
function of first kind:
\begin{equation}\label{rw5c}
K(\vec r,t) = \frac{1}{2\pi}\delta\left(\tau^2\right)-
\frac{1}{4\pi{\ell_{0}}^{2}}
\frac{\ell_{0}}{\tau}
J_{1}\left(\frac{\tau}{\ell_{0}}\right)
\end{equation}
It coincides with the well-known spin zero particle propagator \cite{bogol}.

One can also obtain this result without explicit calculation of
integrals for $I_n$, but solving the integral equation. Indeed, if
we define:
$$
S(\tau)=I_2+I_3(\tau)+I_4(\tau)+\cdots,
$$
then using the formula (\ref{rw5}), we get:
$$
S(\tau)=I_2-\dfrac{1}{\ell_{0}^{2}} \int_{0}^{\tau}
\left[I_2+I_3(\tau')+\cdots\right]\dfrac{(\tau
^{2}-{\tau^{\prime}}^{2})}{2 \tau^{2}}\tau^{\prime}\dd
{\tau^{\prime}},
$$
that is:
\begin{equation}\label{rw5d}
S(\tau)=-\dfrac{1}{8\pi\,{\ell_{0}}^{\!\!2}}-\dfrac{1}{\ell_{0}^{2}}
\int_{0}^{\tau} S(\tau')\dfrac{(\tau ^{2}-{\tau^{\prime}}^{2})}{2
\tau^{2}}\tau^{\prime}\dd {\tau^{\prime}}.
\end{equation}
The solution of this equation (which can be transformed into a
differential equation) is: $$S(\tau)= \frac{-1}{4\pi{\ell_{0}}^{2}}
\frac{\ell_{0}}{\tau} J_{1}\bigl(\frac{\tau}{\ell_{0}}\bigr)$$
which is the same result as in the propagator (\ref{rw5c}).

\section{Mean distance between scatterings}\label{md}
Let us find now the average distance covered by the particle  between
two scatterings.
We repeat for the three dimensions case the same calculation
done in  \cite{Karmanov93} for the one-dimensional case. Namely,  we
will calculate
the number of paths $n_k(t)$ with $k$ events
during  the time $t$. The total number of paths is:
$$n(t)=\sum_k n_k(t).$$
Then we find the propability to have $k$ events for the
time $t$: $$w_k(t) = \dfrac{n_k(t)}{n(t)}.$$ The average number of
events for the time
$t$ is
\begin{equation}\label{eq0}
\bar{k}(t)=\sum_k k w_k(t).
\end{equation}
The time $t$ divided by the number of events for the time $t$, that is,
the ratio $\bar{t}=t/\bar{k}(t)$ for
$t\to\infty$ gives the  average time between two events. Since the speed
between scatterings is always  $c$, in this way we will find the  average
path between events $\bar{r}=c\bar{t}$.

At first we find the number of paths $n_k(t)$ with $k$ events for the time $t$:
\begin{equation}\label{eq1}
n_{k}(t) = \int\delta(t_{1}+t_{2}+\ldots +t_{k}+t_{k+1}-t)\,
\frac{d^{3}r_{1}}{\ell_0^3}\ldots \frac{d^{3}r_{k}}{\ell_0^3}
\frac{d^{3}r_{k+1}}{\ell_0^3}
\end{equation}
where $t_i=r_i/c$.
There are $k+1$ intervals and $k$ scatterings in the points where the intervals
are
linked with each other. Therefore $n_{k}$ is determined by $k+1$ integrations.
The dimensionless integration measure is $\frac{d^{3}r_{i}}{\ell_0^3}$.
Below we put $\ell_0=1$ and restore it in the final formulas.

For $k=0$:
\begin{equation}\label{eq2}
n_{0}(t) = \int\delta(r_{1}-t)\,
d^{3}r_{1}=4\pi \int_0^{\infty}\delta(r_{1}-t)r_{1}^2 dr_{1}=4\pi t^2
\end{equation}
For $k=1$:
\begin{eqnarray}\label{eq3}
n_{1}(t) &=& \int\delta(r_{1}+r_{2}-t)\,
d^{3}r_{1}d^{3}r_{2}
=(4\pi)^2 \int_0^{\infty}\int_0^{\infty}
\delta(r_{1}+r_{2}-t)r_{1}^2 r_{2}^2dr_{1}dr_{2}
\nonumber\\
&=&(4\pi)^2 \int_0^{t}r_{1}^2 (t-r_{1})^2dr_{1}
=(4\pi)^2 \frac{t^5}{30}
\end{eqnarray}
Note that the function $\delta(r_{1}+r_{2}-t)$ does not depend on angles,
therefore the
angle integrations give simply $(4\pi)^2$.

One can easily find the following recurrence formula:
$$
n_{k}(t) =4\pi \int_0^{t}n_{k-1}(t-r_k)r_k^2 dr_k
$$
which is represented in the form:
\begin{equation}\label{eq4}
n_{k}(t) =4\pi \int_0^{t}n_{k-1}(t')(t-t')^2 dt'
\end{equation}

    From the above equations we can see that $n_{k}(t)$ has the form:
\begin{equation}\label{eq5}
n_{k}(t)=c_{k} t^{2+3k}
\end{equation}

    From the recurrence relation (\ref{eq4}) one can derive the
integral equation
for the sum $n(t)$ it gives:
\begin{eqnarray*}
n(t)&=&n_0(t)+n_1(t)+n_2(t)+\ldots
\\
&=&
n_0(t)+ 4\pi \int_0^{t}n_{0}(t')(t-t')^2 dt'
+ 4\pi \int_0^{t}n_{1}(t')(t-t')^2 dt'+\ldots
\\
&=&
n_0(t)+ 4\pi \int_0^{t}[n_0(t')+n_1(t')+n_2(t')+\ldots](t-t')^2 dt'
\end{eqnarray*}
Or:
\begin{equation}\label{eq7}
n(t)=4\pi t^2+ 4\pi\int_0^{t}n(t')(t-t')^2 dt'
\end{equation}

Calculating the third derivative over $t$ of $n(t)$ defined by eq. (\ref{eq7}),
we get the equivalent differential equation:
\begin{equation}\label{eq8}
n'''(t)=8\pi n(t)
\end{equation}
with the initial conditions:
     \begin{equation}\label{eq8a}
n(0)=0,\quad n'(0)=0,\quad n''(0)=8\pi.
     \end{equation}
The solution of eq. (\ref{eq8}) satisfying the initial conditions (\ref{eq8a})
has the form:
\begin{equation}\label{eq8b}
n(t)=\frac{2\pi^{1/3}}{3}\exp(2 \pi^{1/3} t)
-\frac{2\pi^{1/3}}{3}\exp(- \pi^{1/3} t)\left[\cos(\sqrt{3}\pi^{1/3} t)
+\sqrt{3}\sin(\sqrt{3}\pi^{1/3} t)\right],
\end{equation}
that can be checked by the direct substitution.

The decomposition of (\ref{eq8b}) in the Taylor series reads:
$$
n(t)=4\pi t^2+ (4\pi)^2\frac{t^5}{30} +(4\pi)^3\frac{t^8}{5040}+\ldots
$$
that indeed coincides with the sum $n(t)=n_0(t)+n_1(t)+n_2(t)+\ldots$.

Now we calculate $k(t)=\sum_k k n_k(t)$.
Namely, according to eq. (\ref{eq5}), $n(t)$ is given by the sum:
$$
n(t)= \sum_{k=0}^{\infty}  c_{k} t^{2+3k}
$$
and similarly for $k(t)$:
$$
k(t)=\sum_{k=0}^{\infty}  k n_k(t)=\sum_{k=0}^{\infty} k c_{k} t^{2+3k}.
$$
Calculating the first derivative of $n(t)$ over $t$, we get:
$$
n'(t)= \sum_{k=0}^{\infty} (2+3k) c_{k} t^{2+3k-1}=
\frac{2}{t}\sum_{k=0}^{\infty} c_{k}  t^{2+3k}
+\frac{3}{t}\sum_{k=0}^{\infty} k c_{k} t^{2+3k}
=\frac{2}{t}n(t)+\frac{3}{t}k(t).
$$
That is:
\begin{equation}\label{eq9}
k(t)=\frac{1}{3}\left[t n'(t)-2n(t)\right]
\end{equation}
    From (\ref{eq0}) and (\ref{eq9}) we find $\bar{k}(t)$:
\begin{equation}\label{eq10}
\bar{k}(t)=\sum_k k w_k(t)=\frac{\sum_k k n_k(t)}{n(t)}=
\frac{k(t)}{n(t)}=\frac{1}{3}\left[\frac{t n'(t)}{n(t)}-2\right].
\end{equation}
Substituting here, for $t\to \infty$, the leading term of the
expression (\ref{eq8b}) for $n(t)$:
$$
n(t)|_{t\to\infty}\approx\frac{2\pi^{1/3}}{3}\exp(2 \pi^{1/3} t)
$$
we get:
$$
\bar{k}(t)|_{t\to\infty}=\frac{2\pi^{1/3}}{3}t.
$$
For large $t$,  the average number of events linearly increases
with $t$, like for the Poisson distribution.

The average time between events is
$$
\bar{t}=\frac{t}{\bar{k}(t)}=\frac{3}{2\pi^{1/3}}.
$$

    From here we obtain the average distance between two events
$\bar{r}=c\bar{t}$:
\begin{equation}\label{eq11}
\bar{r}=\frac{3}{2\pi^{1/3}} \ell_0.
\end{equation}
Numerically it gives:
$$
\bar{r}=1.02418 \,\ell_0,
$$
a value close to the Compton length.

\section{Conclusions}
In conclusion, the following remarks seem relevant.

     It turned out that we had to suppose that the scattering
     wave v.s. the distance $r$ from the scattering center be decreased
as $1/r$, eq. (\ref{rw1}).
     It looks like the diffusion of a spherical wave which seems  natural,
but, at first glance, it has no any hint to relativity.
However, if, instead of $1/r$, we would take another function of $r$, we
would not reproduce the propagator, but also would loose the relativistic
invariance. In this case we would obtain an expression depending on $t$ and
$r$ taken separately, not in the invariant combination $\tau^2=t^2-r^2$.
This ``no choice'' seems unexpected and intriguing.

      On the other hand, in the stochastic motion of our particle, at each
change of its direction, we supposed that its scattering mimics
resonance S-wave scattering with an amplitude $f=-1$ and a $ \pi/2
$ phase shift. One can say that the "scatterings" which result in
the propagator take place at very small distances, so that higher
partial waves are suppressed.

     Moreover it is interesting to compare the 1-D case with the 3-D one.
     In the 1-D case the average path
between events was found to be exactly the Compton length
$\ell_0$, whereas in this 3-D study it gets
$\dfrac{3}{2\pi^{1/3}}\ell_0 \approx 1.02418 \,\ell_0$. This gives
an example, how a  fundamental characteristic of elementary
particle can be calculated and expressed through simple  numbers,
like $2,3,\pi, \ldots$. Other examples can be found in
\cite{P.Noyes}.

    We have clearly shown that the model is very constrained by our
assumptions. We
think that obtaining the same propagator for the particle as in
standard  quantum
fields theory is not a pure accident. For sure, one should try to find
out physical reasons resulting in the scatterings of the propagating particle
on its path. As already said, some  hypothesis have been advanced in many
papers but it was, at present, beyond the scope of this work.  It would be also
interesting to see how to introduce the spin property in our 3-D model .
\bigskip

{\bf Acknowledgments.}\; One of the authors (V.A.K.) is sincerely
grateful for warm hospitality of the Laboratoire de Physique
Corpusculaire, Universit\'e Blaise Pascal, Clermont-Ferrand,
France, where this work was performed. This work is partially
supported by the French-Russian PICS and RFBR grants Nos. 1172 and
01-02-22002 as well as by the RFBR grant 02-02-16809.

\end{document}